\documentclass[aps,prd,onecolumn,groupedaddress,showpacs,nofootinbib,amssymb]{revtex4}
\usepackage[dvips]{graphicx}
\usepackage{amssymb}
\usepackage{amsmath}
\usepackage{graphicx,,color}
\usepackage{amsfonts}
\usepackage{bm}
\usepackage{cancel}
\usepackage{comment}

\newcommand\be{\begin{equation}}
\newcommand\ee{\end{equation}}

\allowdisplaybreaks[4]

\begin{document}

\tolerance=5000

\title{Rescaled Einstein-Hilbert Gravity from $f(R)$ Gravity: Inflation, Dark Energy and the Swampland Criteria}
\author{V.K.~Oikonomou,$^{1,2}$\,\thanks{v.k.oikonomou1979@gmail.com}}
\affiliation{$^{1)}$ Department of Physics, Aristotle University
of Thessaloniki, Thessaloniki 54124,
Greece\\
$^{2)}$ Laboratory for Theoretical Cosmology, Tomsk State
University of Control Systems and Radioelectronics, 634050 Tomsk,
Russia (TUSUR)}

\tolerance=5000

\begin{abstract}
We consider the scenario that the effective gravitational
Lagrangian of a minimally coupled scalar field at large curvatures
is described by a rescaled Einstein-Hilbert gravity, with the
Ricci scalar being multiplied by a dimensionless parameter
$\alpha$. Such Lagrangian densities might originate for example,
by a class of exponential models of $f(R)$ gravity, in the
presence of a canonical scalar field, for which at early times the
effective Lagrangian of the theory becomes that of a rescaled
canonical scalar field with the Einstein-Hilbert term becoming
$\sim \alpha R$, with $\alpha$ a dimensionless constant, and the
resulting theory is an effective theory of a Jordan frame
$f(R,\phi)$ theory. This rescaled Einstein-Hilbert canonical
scalar field theory at early times has some interesting features,
since it alters the inflationary phenomenology of well-known
scalar field models of inflation, but more importantly, in the
context of this rescaled theory, the Swampland criteria are easily
satisfied, assuming that the scalar field is slowly rolling. We
consider two models of inflation to exemplify our study, a fibre
inflation model and a model that belongs to the general class of
supergravity $\alpha$-attractor models. The inflationary
phenomenology of the models is demonstrated to be viable, and for
the same set of values of the free parameters of each model which
ensure their inflationary viability, all the known Swampland
criteria are satisfied too, and we need to note that we assumed
that the first Swampland criterion is marginally satisfied by the
scalar field, so $\phi\sim M_p$ during the inflationary era.
Finally, we examine the late-time phenomenology of the fibre
inflation potential in the presence of the full $f(R)$ gravity,
and we demonstrate that the resulting model produces a viable dark
energy era, which resembles the $\Lambda$-Cold-Dark-Mater model.
Thus in the modified gravity model we present, the Universe is
described by a rescaled Einstein-Hilbert gravity at early times,
hence in some sense the modified gravity effect is minimal
primordially, and the scalar field controls mainly the dynamics
with a rescaled Ricci scalar gravity. However the effect of $f(R)$
gravity becomes stronger at late-times, where it controls the
dynamics, synergistically with the scalar field.
\end{abstract}


\maketitle

\section{Introduction}

The main mysteries in cosmology are related strongly to the
accelerating eras of our Universe, namely the inflationary era and
the dark energy era of our Universe. Both these two eras are
currently under theoretical and observational investigation, and
to date, no concrete model exists that describes these two eras,
although many viable candidate theories appear in the literature.
With regard to inflation, from the observational point of view,
the latest Planck data \cite{Akrami:2018odb} have put strong
constraints on the observational indices of inflation, however, to
date there is no sign for the presence of $B$-modes (curl), which
would definitely verify the occurrence of an inflationary era. It
is useful to note that the B-mode polarization is due to two
reasons, firstly from primordial tensor perturbations on large
angular scales (so at low multipoles of the Cosmic Microwave
Background $l \leq 10$), or at late times due to gravitational
lensing conversion of E-modes to B-modes which occur for small
angular scales \cite{Denissenya:2018mqs}. However, not only
inflationary theories lead to a primordial tensor perturbations
spectrum, but also alternatives to inflation do, which typically
produce more B-mode polarization (the exception being the initial
version of the Ekpyrotic scenario). With regard to the dark energy
era, this was firstly observed in the late 90's
\cite{Riess:1998cb}, based on observations of Type IA supernovas
which are white dwarfs standard candles. However, to date, only
constraints on several physical quantities related to dark energy
are obtained, and although many candidate theories can potentially
describe the dark energy era, no definitive answer is given for
the question what is the nature of dark energy. Also we need to
note that there exist studies in the literature that the dark
energy era can be combined with a viable bouncing era
\cite{Odintsov:2020vjb}, with the bouncing cosmology scenario
being an interesting alternative to the inflationary scenario the
bouncing cosmology scenario
\cite{Brandenberger:2012zb,Brandenberger:2016vhg,Battefeld:2014uga,Novello:2008ra,Cai:2014bea,deHaro:2015wda,Lehners:2011kr,Lehners:2008vx,Cheung:2016wik,Cai:2016hea}.

For inflation, the usual description uses the inflaton scalar
field
\cite{Linde:2007fr,Gorbunov:2011zzc,Lyth:1998xn,Martin:2018ycu},
and from a string theory point of view, scalar fields seem to play
a prominent role in nature. In fact, the Standard Model of
particle physics is built upon the Higgs scalar which gives mass
to all massive particles. On the other hand, since inflation is
marginally chronically near to the quantum gravity era, it is
possible that higher order curvature terms might appear in the
inflationary Lagrangian. Modified gravity theory, describes such
models that contain higher order curvature terms or even string
motivated terms
\cite{reviews1,reviews2,reviews3,reviews4,reviews5,reviews6}, and
is known to describe inflation and dark energy under the same
theoretical framework, with the unification of inflation and dark
energy being firstly proposed in Ref. \cite{Nojiri:2003ft}, and
later further developed in Refs.
\cite{Nojiri:2007as,Nojiri:2007cq,Cognola:2007zu,Nojiri:2006gh,Appleby:2007vb,Elizalde:2010ts,Odintsov:2020nwm}.
Motivated by this, in this paper we shall consider a class of
exponential models of $f(R)$ gravity, firstly considered in Ref.
\cite{Oikonomou:2020qah}, in the presence of a canonical scalar
field with scalar potential. The effective Lagrangian during the
large curvature regime which is realized during the inflationary
era, at leading order becomes that of a pure canonical scalar
field in the presence of a rescaled Einstein-Hilbert term of the
form $\sim \alpha R$, with $\alpha$ being a dimensionless
constant. The resulting rescaled Einstein-Hilbert theory can alter
the phenomenology of several known inflationary models, however,
in a positive way, since the phenomenological viability of the
models also occurs for the rescaled theory. However, the attribute
of this rescaled theory is that all the swampland criteria might
be satisfied for the same set of free parameter values which
render the inflationary theory phenomenologically viable. The
Swampland conjectures, firstly appeared in Refs.
\cite{Vafa:2005ui,Ooguri:2006in} and were further studied in Refs.
\cite{Palti:2020qlc,Mizuno:2019bxy,Brandenberger:2020oav,Blumenhagen:2019vgj,Wang:2019eym,Benetti:2019smr,Palti:2019pca,Cai:2018ebs,Akrami:2018ylq,Mizuno:2019pcm,Aragam:2019khr,Brahma:2019mdd,Mukhopadhyay:2019cai,Brahma:2019kch,Haque:2019prw,Heckman:2019dsj,Acharya:2018deu,Elizalde:2018dvw,Cheong:2018udx,Heckman:2018mxl,Kinney:2018nny,Garg:2018reu,Lin:2018rnx,Park:2018fuj,Olguin-Tejo:2018pfq,Fukuda:2018haz,Wang:2018kly,Ooguri:2018wrx,Matsui:2018xwa,Obied:2018sgi,Agrawal:2018own,Murayama:2018lie,Marsh:2018kub,Storm:2020gtv,Trivedi:2020wxf,Sharma:2020wba,Odintsov:2020zkl,Mohammadi:2020twg,Trivedi:2020xlh,Han:2018yrk,Achucarro:2018vey,Akrami:2020zfz},
see also \cite{Colgain:2018wgk,Colgain:2019joh,Banerjee:2020xcn}
for the $H_0$ tension problem relation with the Swampland
criteria. Impose some restrictions in scalar theories, since the
slow-roll conditions might be incompatible with the Swampland
criteria. In our case, the $\alpha$-rescaled theory might lead to
both the phenomenological inflationary viability of the scalar
models and at the same time the theory is compatible with the
Swampland criteria. This result is based on an initial $f(R)$
gravity theory which at leading order in the high curvature
regime, yields the rescaled Einstein-Hilbert term. In order to
exemplify our claims, we study two well-known models of inflation,
fibre inflation \cite{Stewart:1994ts,Cicoli:2008gp} and a
supergravity $\alpha$-attractor model \cite{Carrasco:2015rva}. As
we demonstrate, both the models yield a viable inflationary
phenomenology and are also compatible with the Swampland criteria,
for the same set of values of the free parameters that render the
inflationary theory viable.

Apart from the inflationary era, we also discuss the dark energy
era of the combined $f(R)$ scalar field model. We choose the fibre
inflation model in the presence of the full $f(R)$ gravity, to
quantify our study, and by reexpressing the Friedmann equation in
terms of the redshift and of a suitable dark energy statefinder
quantity, we numerically solve the resulting Friedmann equation.
As we demonstrate, the model is compatible with the Planck data on
the cosmological parameters for some physical quantities of great
phenomenological interest, and also when statefinders are
considered, the model resembles the behavior of the
$\Lambda$-Cold-Dark-Matter ($\Lambda$CDM) model. An interesting
feature of the model is that at late-times the dark energy era is
not free from dark energy oscillations. This result is intriguing,
since a similar exponential $f(R)$ gravity model does not yield
dark energy oscillations at late times. Thus we conclude that the
dark energy oscillations feature is somewhat model dependent.

\section{The $f(R)$ Gravity Scalar Field Model, Inflationary Effective Theory and the Swampland Criteria}

The $f(R)$ gravity scalar field model action has the following
form,
\begin{equation}
\label{action} \centering
\mathcal{S}=\int{d^4x\sqrt{-g}\left(\frac{f(R)}{2\kappa^2}-\frac{1}{2}\partial_\mu\phi\partial^\mu\phi-V(\phi)\right)}\,
,
\end{equation}
with $\kappa=\frac{1}{M_p}$ and $M_P$ denotes is the reduced
Planck mass. The $f(R)$ gravity will be assumed to have the form,
\begin{equation}\label{frini}
f(R)=R-\gamma  \lambda  \Lambda -\lambda  R \exp
\left(-\frac{\gamma  \Lambda }{R}\right)-\frac{\Lambda
\left(\frac{R}{m_s^2}\right)^{\delta }}{\zeta }\, ,
\end{equation}
where $\Lambda$ will be assumed to take values of the order of the
cosmological constant at present time, so it has dimensions
eV$^2$, while the parameter $m_s$ is defined to be
$m_s^2=\frac{\kappa^2\rho_m^{(0)}}{3}$, with $\rho_m^{(0)}$ being
the energy density of cold dark matter at present day. Also, the
parameters $\lambda$, $\gamma$, $\delta$ and $\zeta$ are
dimensionless parameters to be defined during the inflationary
era. The only constraint we assume on the value of $\delta$ is
$\delta<0.01$ in order for this term to be subleading during the
inflationary era. In fact the motivation for adding this term comes
from the dark energy era study, in order to produce more smooth
results.

By assuming a flat Friedmann-Robertson-Walker (FRW) metric,
\begin{equation}
\label{metric} \centering
ds^2=-dt^2+a(t)^2\sum_{i=1}^{3}{(dx^{i})^2}\, ,
\end{equation}
the field equations for the theory (\ref{action}) are,
\begin{equation}
\label{motion1} \centering
3H^2f_R=\frac{Rf_R-f}{2}-3H\dot{f}_R+\kappa^2\left(
\frac{1}{2}\dot\phi^2+V\right)\, ,
\end{equation}
\begin{equation}
\label{motion2} \centering -2\dot
Hf_R=\kappa^2\dot\phi^2+\ddot{f}_R-H\dot{f}_R\, ,
\end{equation}
\begin{equation}
\label{motion3} \centering \ddot{\phi}+3H\dot{\phi}+V'=0\, .
\end{equation}
where $f_R=\frac{d f}{d R}$. Let us consider how the $f(R)$
gravity model is approximated during the early-time era, and since
the curvature even for the low-scale inflation is of the order
$\sim H_I^2$, with $H_I\sim 10^{13}$GeV, this means that during
the inflationary era, the curvature satisfies $R\gg \Lambda
\gamma$. Thus the exponential term in the $f(R)$ gravity of Eq.
(\ref{frini}) during the large curvature regime can be
approximated as follows,
\begin{equation}\label{expapprox}
\lambda  R \exp \left(-\frac{\gamma  \Lambda }{R}\right)\simeq
-\gamma \lambda  \Lambda -\frac{\gamma ^3 \lambda \Lambda^3}{6
R^2}+\frac{\gamma ^2 \lambda  \Lambda ^2}{2 R}+\lambda  R\, ,
\end{equation}
thus the effective action during the inflationary era is at
leading order in $R$,
\begin{equation}\label{effectiveaction}
\mathcal{S}=\int
d^4x\sqrt{-g}\left(\frac{1}{2\kappa^2}\left(\alpha R+ \frac{\gamma
^3 \lambda \Lambda ^3}{6 R^2}-\frac{\gamma ^2 \lambda \Lambda
^2}{2 R}-\frac{\Lambda}{\zeta
}\left(\frac{R}{m_s^2}\right)^{\delta
}+\mathcal{O}(1/R^3)+...\right)-\frac{1}{2}\partial_\mu\phi\partial^\mu\phi-V(\phi)\right)\,
,
\end{equation}
where we have set $\alpha=1-\lambda$. Thus by taking into account
that the parameter $\delta$ satisfies $\delta<0.01$, the effective
action corresponding to the large curvature regime of the action
(\ref{action}) is basically a rescaled Einstein-Hilbert canonical
scalar field theory. It is notable that the rescaled
Einstein-Hilbert theory is a Jordan frame theory, the Einstein
frame counterpart of which would contain two scalar fields, thus
the action (\ref{effectiveaction}) is an effective theory in
Jordan frame, and being such we cannot make the correspondence
with the Einstein frame, because we should take into account the
other $F(R)$ gravity terms into account.

Let us now consider the phenomenology of the rescaled
Einstein-Hilbert scalar model of Eq. (\ref{effectiveaction}).
Firstly, for the effective action (\ref{effectiveaction}), the
field equations of the effective theory (\ref{effectiveaction}) at
leading order read,
\begin{equation}
\label{motion1a} \centering 3H^2\alpha\simeq \kappa^2\left(
\frac{1}{2}\dot\phi^2+V\right)\, ,
\end{equation}
\begin{equation}
\label{motion2a} \centering -2\dot H\alpha\simeq
\kappa^2\dot\phi^2\, ,
\end{equation}
\begin{equation}
\label{motion3a} \centering \ddot{\phi}+3H\dot{\phi}+V'=0\, .
\end{equation}
The slow-roll indices for the above theory are defined as follows,
\cite{Hwang:2005hb},
\begin{align}\label{slowrollindicesdef}
& \epsilon_1=-\frac{\dot H}{H^2}, \\ \notag &
\epsilon_2=\frac{\ddot\phi}{H\dot\phi},
\end{align}
and the spectral index of the primordial curvature scalar
perturbations is,
\begin{equation}\label{spectralindexdef}
n_s=1-4 \epsilon_1-2\epsilon_2\, .
\end{equation}
Let us calculate in detail the two slow-roll indices by taking
into account the slow-roll assumptions $\ddot{\phi}\ll H
\dot{\phi}$, and $\frac{\dot{\phi}^2}{2}\ll V$, hence the scalar
field equation reads,
\begin{equation}
\label{motion3appr} \centering 3H\dot{\phi}+V'\simeq 0\, ,
\end{equation}
while the Friedmann equation reads,
\begin{equation}
\label{motion1appr} \centering 3H^2\alpha\simeq \kappa^2V\, ,
\end{equation}
thus, the first slow-roll index $\epsilon_1$ in view of Eqs.
(\ref{motion3appr}) and (\ref{motion1appr}) becomes,
\begin{equation}\label{firstslowrollapproxi}
\epsilon_1=\frac{\alpha}{2\kappa^2}\left(\frac{V'}{V}
\right)^2=\alpha \epsilon\, ,
\end{equation}
where $\epsilon$ is the usual first slow-roll index of the
canonical scalar theory
\begin{equation}\label{slowrollcanonical}
\epsilon=\frac{1}{2\kappa^2}\left(\frac{V'}{V} \right)^2\, .
\end{equation}
In addition, the second slow-roll index $\epsilon_2$ is formally
written,
\begin{equation}\label{secondslowrollapproxi}
\epsilon_2=\frac{\ddot{\phi}}{H\dot{\phi}}\simeq
\partial_t\left(-\frac{V'}{3H}\right)/(H\dot{\phi})\simeq -\alpha \eta+\epsilon_1\, ,
\end{equation}
where $\eta$ is the second slow-roll index for the canonical
scalar theory, defined as,
\begin{equation}\label{secondslowrollindex}
\eta=\frac{1}{\kappa^2}\frac{V''}{V}\, .
\end{equation}
In view of Eqs. (\ref{firstslowrollapproxi}) and
(\ref{secondslowrollapproxi}), the spectral index
(\ref{spectralindexdef}) reads,
\begin{equation}\label{spectralindexfinal}
n_s\simeq 1-6\alpha \epsilon+2\alpha \eta\, ,
\end{equation}
thus the difference with the canonical Einstein-Hilbert theory is
the presence of the parameter $\alpha$. Let us now turn our focus
to the tensor-to-scalar ratio, which is defined as follows
\cite{reviews1},
\begin{equation}\label{tensordef}
r=\frac{8\kappa^2Q_s}{f_R}\, ,
\end{equation}
with $Q_s=\frac{\dot{\phi}^2}{H^2}$. By using the Raychaudhuri
equation (\ref{motion2a}) we have
$Q_s=-\frac{2\dot{H}\alpha}{\kappa^2H^2}$, hence the resulting
expression for the tensor to scalar ratio is,
\begin{equation}\label{final}
r\simeq 16 \epsilon_1=16 \alpha \epsilon\, .
\end{equation}
Let us now investigate how the $e$-foldings number becomes in
terms of the scalar field. We have,
\begin{equation}\label{efoldingsnew}
N=\int_{t_i}^{t_f}H
dt=\frac{\kappa^2}{\alpha}\int_{\phi_f}^{\phi_i}\frac{V}{V'}
d\phi\, .
\end{equation}
By using the expressions (\ref{spectralindexfinal}), (\ref{final})
and (\ref{efoldingsnew}), one may study the inflationary
phenomenology following the well known procedure for scalar field
inflation. The final value of the scalar field at the end of
inflation is calculated by solving the condition
$\epsilon_1(\phi_f)=1$ with respect to the scalar field, and thus
by integrating (\ref{efoldingsnew}) one may determine the value of
the scalar field at the first horizon crossing, namely $\phi_i$.
Accordingly, by calculating $n_s$ and $r$ for the value of the
scalar field at the first horizon crossing, one may have analytic
expressions for both $n_s$ and $r$ as functions of the
$e$-foldings and the free parameters of the scalar field model. In
the following we shall use the above formalism in order to examine
the phenomenological viability of two scalar field models. Before
proceeding to this, let us recall the Swampland criteria
\cite{Vafa:2005ui,Ooguri:2006in,Palti:2020qlc,Mizuno:2019bxy,Brandenberger:2020oav,Blumenhagen:2019vgj,Wang:2019eym,Benetti:2019smr,Palti:2019pca,Cai:2018ebs,Akrami:2018ylq,Mizuno:2019pcm,Aragam:2019khr,Brahma:2019mdd,Mukhopadhyay:2019cai,Brahma:2019kch,Haque:2019prw,Heckman:2019dsj,Acharya:2018deu,Elizalde:2018dvw,Cheong:2018udx,Heckman:2018mxl,Kinney:2018nny,Garg:2018reu,Lin:2018rnx,Park:2018fuj,Olguin-Tejo:2018pfq,Fukuda:2018haz,Wang:2018kly,Ooguri:2018wrx,Matsui:2018xwa,Obied:2018sgi,Agrawal:2018own,Murayama:2018lie,Marsh:2018kub,Storm:2020gtv,Trivedi:2020wxf,Sharma:2020wba,Odintsov:2020zkl,Mohammadi:2020twg,Trivedi:2020xlh,Colgain:2018wgk,Colgain:2019joh,Banerjee:2020xcn},
which are the following assuming reduced Planck units
($\kappa^2=1$),
\begin{enumerate}
    \item $\Delta \phi \leq d$, with $d\sim \mathcal{O}(1)$ in
    reduced Planck units.
    \item $\Big{|}\frac{V'}{V} \Big{|}\geq \gamma$, with $\gamma\sim \mathcal{O}(1)$ in
    reduced Planck units.
    \item $-\frac{V''}{V}\geq 1$
\end{enumerate}
The Swampland criteria for the normal Einstein-Hilbert canonical
scalar theory is the fact that the first slow-roll index must
satisfy,
\begin{equation}\label{firstswampclass}
\Big{|}\frac{V'}{V} \Big{|}=\sqrt{2\epsilon}\geq 1\, ,
\end{equation}
hence the tensor-to-scalar ratio of the ordinary canonical scalar
theory, which is $r=16\epsilon$, must be larger than unity in
reduced Planck units, so this puts the inflationary phenomenology
of scalar models in peril. In addition, the spectral index of the
ordinary canonical scalar theory would be $n_s=1-6\epsilon+2\eta$,
so if $\epsilon\sim \mathcal{O}(1)$, it might be difficult to
obtain a nearly scale invariant spectrum with $n_s\sim
\mathcal{O}(1)$.

Let us discuss how the rescaled Einstein-Hilbert scalar theory
offers a possible remedy for the Swampland issues of the pure
scalar theory, and at the same time, a viable phenomenology may be
obtained by the same theory. In the rescaled Einstein-Hilbert
gravity, the slow-roll conditions for the slow-roll indices are
quantified not by the conditions $\epsilon\ll 1$ and $\eta \ll 1$,
but by the conditions $\epsilon_1\ll 1$ and $\epsilon_2\ll 1$.
Hence, it is obvious that the Swampland conditions can be
satisfied even if $\epsilon\sim \mathcal{O}(1)$, since it is
possible to achieve $\epsilon_1\ll 1$ and $\epsilon_2\ll 1$, by
choosing the parameter $\alpha$ to be small enough in order to
guarantee that the Swampland inequalities are satisfied. To be
more precise, due to the fact that $\epsilon_1=\alpha \epsilon$
even if $\epsilon>1$, if we choose $\alpha$ appropriately, we can
achieve $r=16\alpha \epsilon \ll 1$. Also the spectral index in
the rescaled Einstein-Hilbert gravity is $n_s=1-6 \alpha
\epsilon+2\alpha \eta$, can be very close to unity and compatible
with the Planck data, if the parameter $\alpha$ is appropriately
chosen. Thus in principle it is possible to realize a viable
slow-roll inflationary era, and at the same time to satisfy the
Swampland criteria for the rescaled Einstein-Hilbert gravity
framework. We shall explicitly demonstrate this in the next
section, by using two well-known inflationary models.

Before proceeding, an important comment is in order regarding the
whole $a$-rescaled theoretical framework. On a nutshell, the
effect is seen only at the equation of motion level, and not at
the initial action (\ref{action}), which is used in order to
extract the Swampland criteria. In fact the effective action
(\ref{effectiveaction}) is a virtual action, not a true action
describing the system. It is a virtual action which may produce
the slow-roll equations of motion at leading order. To be
specific, the scalar field present in the initial action, is some
remnant of an effectively massless mode at Planck scale, which
gained mass when the original M-theory was compactified in four
dimensions. The Swampland criteria and the corresponding de Sitter
condition applies to the original 11-dimensional M-theory or even
the 10-dimensional supersymmetric M-theory. This de Sitter
condition is violated at Planck scale, but it is satisfied at the
compactification scale, and during the inflationary era, where
$\phi\sim \mathcal{O}(1)$ in reduced Planck units. The de Sitter
criterion then should constrain the pre-inflationary era and the
inflationary era itself. Our approach in this paper however, is
mainly based on leading order approximations during the slow-roll
era at the equations of motion level. We assume that the starting
action (\ref{action}) is the action for which the Swampland
criteria apply. The rescaled action (\ref{effectiveaction}) is a
virtual effective action which corresponds to the leading order
simplification of the equations of motion, assuming also a
slow-roll evolution. This virtual effective action is the action
that can describe the dynamical evolution of the cosmological
system at leading order, at the level of equations of motion. The
action that would directly extract the slow-roll leading order
equations of motion. Thus the Swampland criteria are not derived
by this effective action, which as we mentioned is a virtual
action, not a real action. Hence, the de Sitter Swampland
criterion derived by the initial action does not contain the
parameter $a$, which enters virtually only the equations of
motion, when leading order curvature terms are kept and in
addition the slow-roll conditions apply. This is similar to the
effect of the slow-roll approximation itself, one could derive an
effective action which could result to the slow-roll approximated
equations of motion, but this effective action by no means is not
the original action of the cosmological system, it is just a
virtual action. Thus, the de Sitter condition constrains directly
$\nabla V/V$, and via the equations of motion this de Sitter
condition can be satisfied. The original action constrains $\nabla
V/V\geq 1$ in reduced Planck units, and for a single scalar field
theory, this would mean that the first slow-roll index
$\epsilon_1=-\frac{\dot{H}}{H^2}$, which under the slow-roll
approximation becomes
$\epsilon_1=\frac{1}{2\kappa^2}\left(\frac{V'}{V} \right)^2$ would
have a direct problem due to the fact that the Swampland de Sitter
criterion implies $V'/V\geq 1$. However, in our case, he original
action (\ref{action}) indicates that $V'/V\geq 1$, however at the
equations of motion level, the first slow-roll index $\epsilon_1$
under the slow-roll approximation and by keeping leading order
curvature terms is explicitly,
\begin{equation}\label{newslowroll}
\epsilon_1=-\frac{\dot{H}}{H^2}\simeq
\frac{\alpha}{2\kappa^2}\left(\frac{V'}{V} \right)^2\, .
\end{equation}
This is the new feature of our framework, the Swampland criterion
is not affected since it applies to the original action
(\ref{action}), or even the Einstein frame transformed one, where
two scalar fields would be present, and although $V'/V$ is
constrained, the first slow-roll index of the theory is not
affected, since it is given by
$\epsilon_1=-\frac{\dot{H}}{H^2}\simeq
\frac{\alpha}{2\kappa^2}\left(\frac{V'}{V} \right)^2$. Thus the
parameter $a$ does not enter the de Sitter Swampland condition.
Hence dynamically, at the equations of motion level, the Swampland
de Sitter condition does not affect the slow-roll dynamics of the
theory, for the rescaled theory we developed, due to the presence
of the parameter $a$. In the single scalar field case, the de
Sitter condition destroys the slow-roll era. In the rescaled
theory we developed, the slow-roll condition is not affected at
all by the Swampland condition. Hence, our approach is generated
by the Jordan frame slow-roll dynamics by keeping leading order
curvature terms. The main result of our works thus, is that the de
Sitter Swampland criterion does not directly affect the slow-roll
condition that the slow-roll index $\epsilon_1$ must satisfy, due
to the presence of the parameter $a$. This is an  equation of
motion effect, generated by the slow-roll conditions and by
keeping leading order curvature terms, and it is not related to
any modified action. The action (\ref{effectiveaction}) is a
virtual action that would generate the equations of motion after
the slow-roll conditions are imposed and if leading order
curvature terms are kept. It is a fictitious virtual action, not a
true action for the system, thus the Swampland criteria are not
derived by it, but from the initial action (\ref{action}).

\section{Inflationary Phenomenology and Swampland Criteria for a Supergravity $\alpha$-Attractor Model}

\begin{figure}
\centering
\includegraphics[width=18pc]{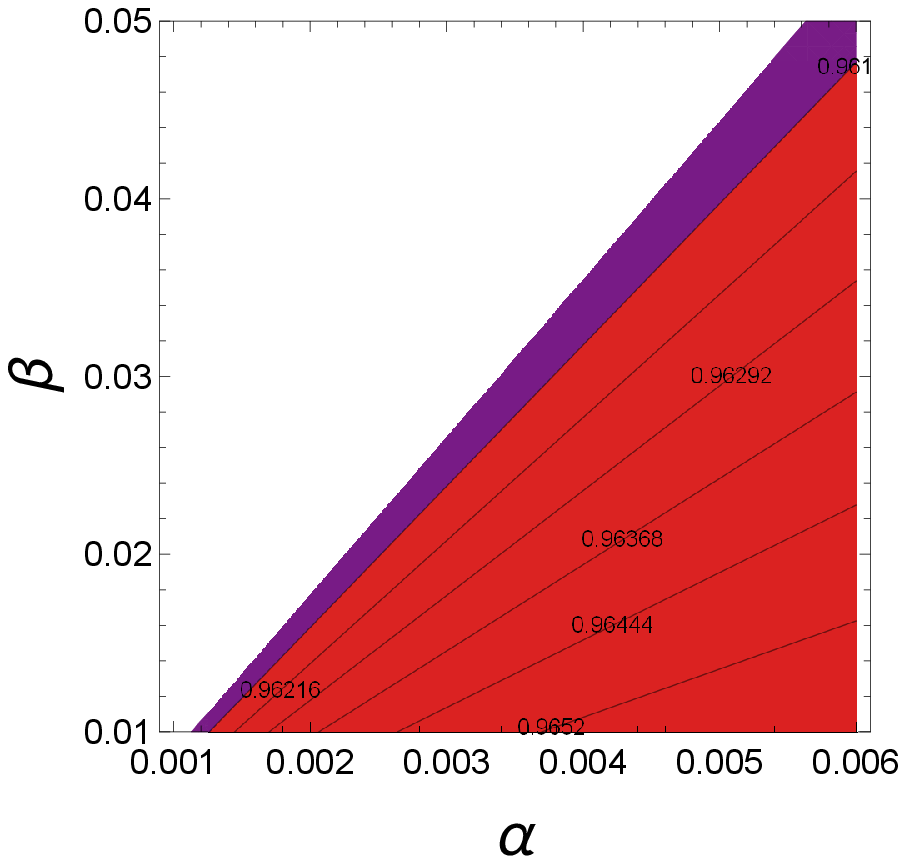}
\includegraphics[width=18pc]{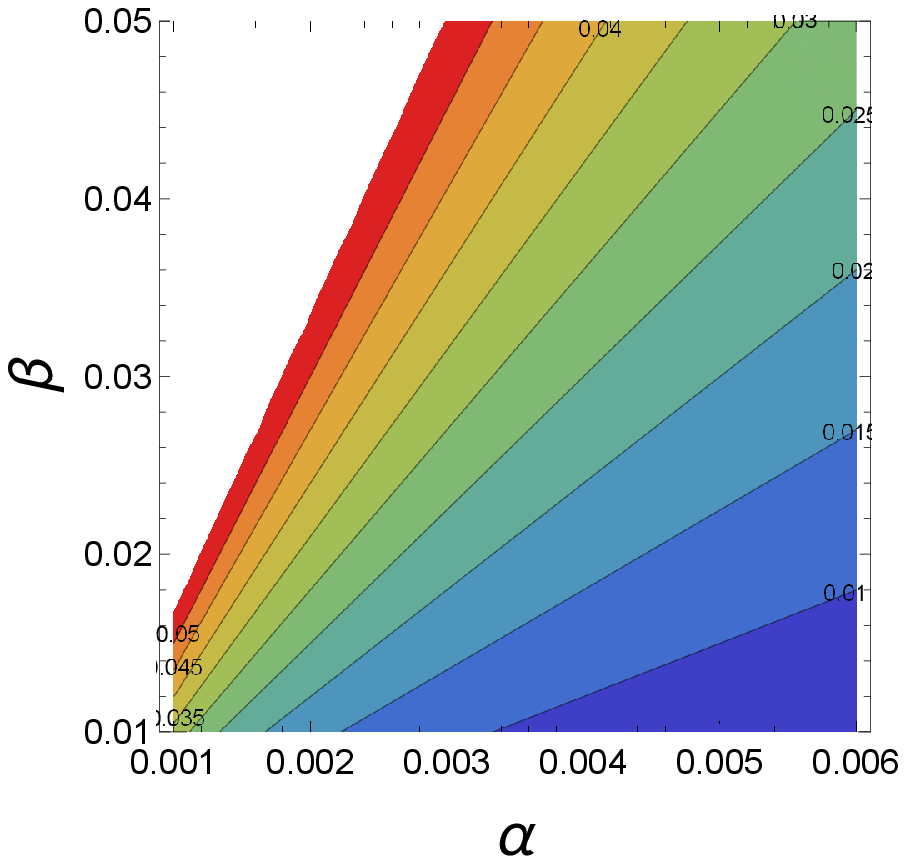}
\caption{Contour plots for the spectral index of primordial scalar
curvature perturbations $n_s$ (left plot) and the tensor-to-scalar
ratio $r$ (right plot), for $\beta=[0.01,0.05]$,
$\alpha=[0,0.006]$ and $N=60$.}\label{plot1}
\end{figure}
\begin{figure}
\centering
\includegraphics[width=18pc]{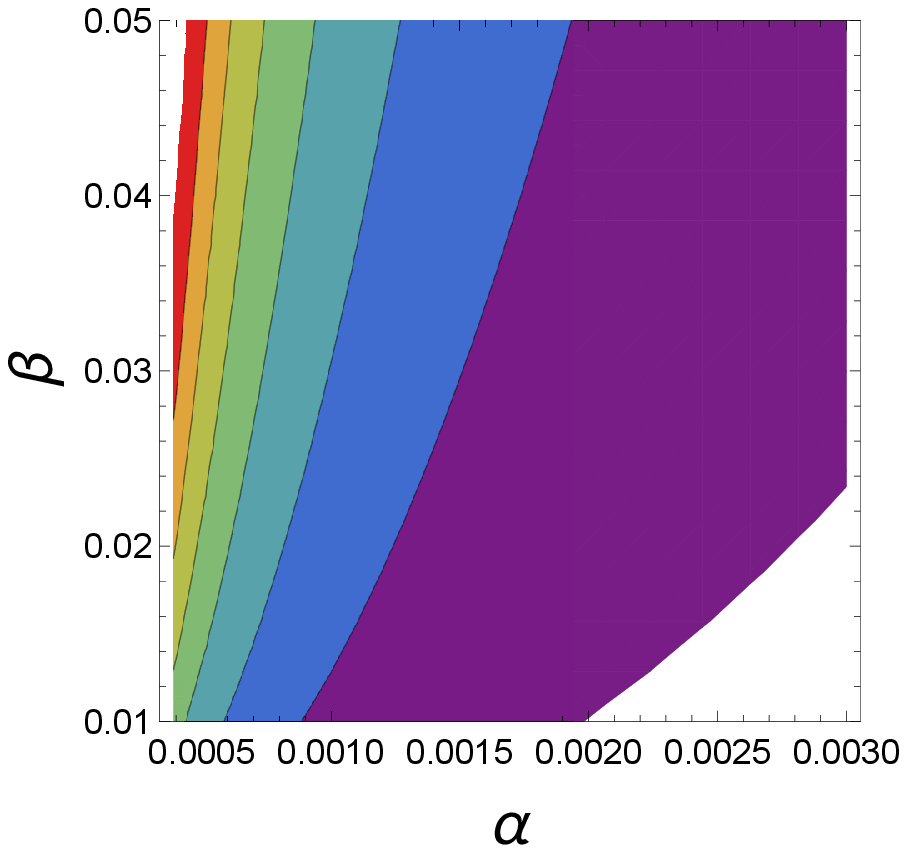}
\includegraphics[width=18pc]{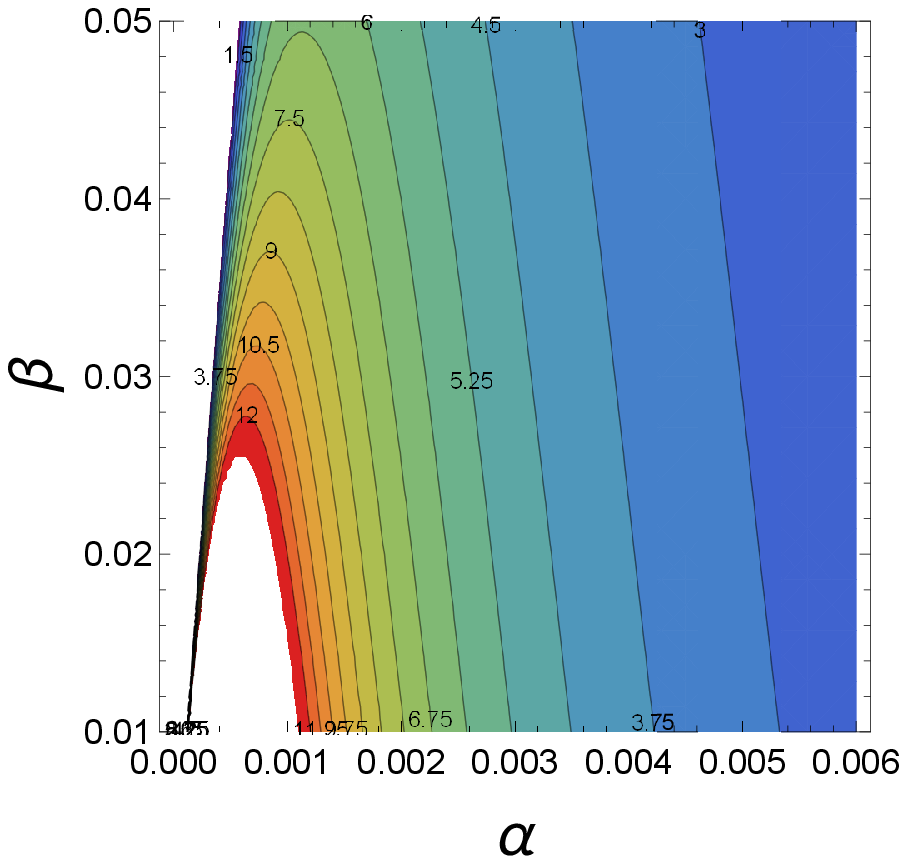}
\caption{Second and third Swampland criteria analysis. Contour
plots of $\frac{V'(\phi)}{V(\phi)}$ (left plot) and of
$-\frac{V''(\phi)}{V(\phi)}$ (right plot), evaluated at the first
horizon crossing for the ranges of values of the free parameters,
$\beta=[0.01,0.05]$ and $\alpha=[0,0.006]$ and for $N=60$, in
reduced Planck units.}\label{plot2}
\end{figure}

Let us start our quantitative study of inflation for the rescaled
Einstein-Hilbert gravity, by analyzing a supergravity
$\alpha$-attractor model, for which the scalar potential is
\cite{Carrasco:2015rva},
\begin{equation}\label{attractor}
V(\phi)=V_0^4 \left(1-e^{-\sqrt{\frac{2}{3 \beta }} \kappa  \phi
}\right)^2\, ,
\end{equation}
where $V_0$ is a constant with mass dimensions $[V_0]=m$ and
$\beta$ is a dimensionless constant. The slow-roll index
$\epsilon_1$ is easily found,
\begin{equation}\label{epsilonalphaatr}
\epsilon_1=\frac{4 \alpha }{3 \beta  \left(e^{\sqrt{\frac{2}{3}}
\sqrt{\frac{1}{\beta }} \kappa  \phi }-1\right)^2}\, ,
\end{equation}
and by solving the equation $\epsilon_1(\phi_f)=1$, we obtain the
value of the scalar field at the end of the inflationary era,
which is,
\begin{equation}\label{phifinal1}
\phi_f=\frac{\sqrt{\frac{3}{2}} \log \left(\frac{2 \sqrt{3}
\sqrt{\alpha }+3 \sqrt{\beta }}{3 \sqrt{\beta
}}\right)}{\sqrt{\frac{1}{\beta }} \kappa }\, .
\end{equation}
Accordingly, by integrating (\ref{efoldingsnew}), we obtain the
value of the scalar field at first horizon crossing,
\begin{equation}\label{phii1}
\phi_i=\frac{\sqrt{\frac{3}{2}} \log \left(\frac{2 \sqrt{3}
\sqrt{\alpha } \sqrt{\beta }+3 \beta +4 \alpha  N}{3 \beta
}\right)}{\sqrt{\frac{1}{\beta }} \kappa }\, .
\end{equation}
Hence, the spectral index of the primordial scalar curvature
perturbations, and the tensor-to-scalar ratio can be evaluated at
leading order in $1/N$, and these are,
\begin{equation}\label{nsalphaattractor}
n_s\simeq 1-\frac{2}{N}+\frac{\sqrt{3} \sqrt{\beta }}{\sqrt{\alpha
} N^2}-\frac{3 \beta }{\alpha  N^2}\, ,
\end{equation}
\begin{equation}\label{tensoralphaattr}
r\simeq \frac{12 \beta }{\alpha  N^2}\, .
\end{equation}
The phenomenological viability of the model can be achieved for a
large range of the free parameters $\alpha$ and $\beta$, for
$N\sim 60$, however we shall choose both $\alpha$ and $\beta$ in
such a way so both the inflationary phenomenology viability and
the compliance with the Swampland conjectures is simultaneously
achieved. In Fig. \ref{plot1} we present the contour plots for the
spectral index of primordial scalar curvature perturbations values
(left plot) and the tensor-to-scalar ratio (right plot), for
$\beta=[0.01,0.05]$ and $\alpha=[0,0.006]$ and $N=60$. Recall that
the latest Planck constraints indicate that \cite{Akrami:2018odb},
\begin{equation}\label{planck1}
n_{s} = 0.9649 \pm 0.0042, \,\,\, r < 0.056 \, ,
\end{equation}
so in both the plots of Fig. \ref{plot1}, we kept values in the
plot range that satisfy the latest Planck constraints, that is
$n_s=[0.9607, 0.9691]$ and $r=[0,0.056]$. As it can be seen, the
simultaneous compatibility of both the spectral index $n_s$ and of
the tensor-to-scalar ratio $r$ is achieved for this range of
values of $\alpha$ and $\beta$. A numerical example would also be
useful here, so for $\alpha=0.004$ and $\beta=0.02$ and for
$N=60$, we get $n_s=0.963576$ and $r=0.0166667$, which are of
course compatible with the latest Planck data (\ref{planck1}). Now
let us turn our focus on the Swampland criteria, and in Fig.
\ref{plot2} we plotted the contour plots of
$\frac{V'(\phi)}{V(\phi)}$ (left plot) and of
$-\frac{V''(\phi)}{V(\phi)}$ (right plot), evaluated at the first
horizon crossing (which is the relevant value of the scalar field
for inflationary phenomenology) for the same range of values of
the free parameters $\alpha$ and $\beta$, that is for,
$\beta=[0.01,0.05]$ and $\alpha=[0,0.006]$ and for $N=60$, in
reduced Planck units. In the left plot we kept the plot values of
$\frac{V'(\phi)}{V(\phi)}$ to be in the range
$\frac{V'(\phi)}{V(\phi)}=[1,9]$ while for
$\frac{V''(\phi)}{V(\phi)}=[1,13]$, so both the quantities
$\frac{V'(\phi)}{V(\phi)}$ and of $-\frac{V''(\phi)}{V(\phi)}$
take values well above unity in reduced Planck units.
Hence, for the $\alpha$-attractor $E$-model, the last two
Swampland criteria are satisfied. Finally, let us note that for
all the analysis performed above, we assumed that the first
Swampland criterion marginally holds true, so during inflation,
the scalar field takes values $\Delta \phi \sim M_p$, or $\Delta
\phi \sim 1$ in reduced Planck units.

\section{Inflationary Phenomenology and Swampland Criteria for a Fibre Inflation Model}

\begin{figure}
\centering
\includegraphics[width=18pc]{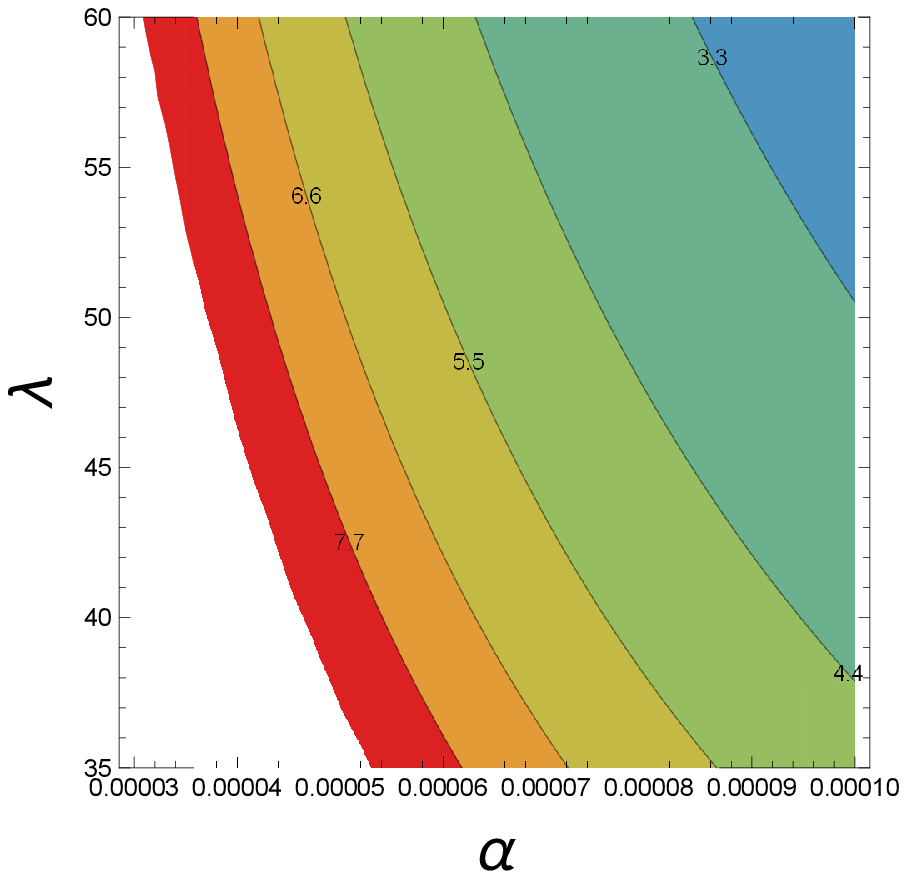}
\includegraphics[width=18pc]{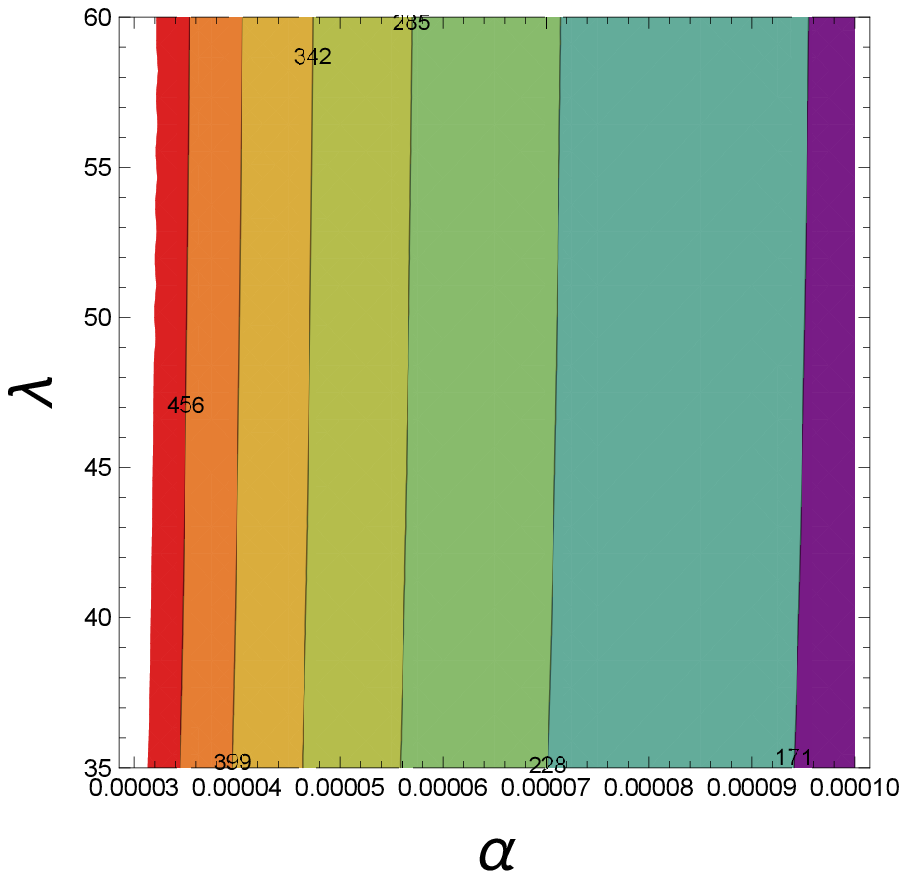}
\caption{Second and third Swampland criteria analysis for the
Fibre inflation model. Contour plots of $\frac{V'(\phi)}{V(\phi)}$
(left plot) and of $-\frac{V''(\phi)}{V(\phi)}$ (right plot),
evaluated at the first horizon crossing for the ranges of values
of the free parameters $\alpha$ and $\lambda$, $\lambda=[35,60]$
and $\alpha=[0.00003, 0.0001]$ and for $N=60$, in reduced Planck
units.}\label{plot4}
\end{figure}

Let us now consider another model of phenomenological interest,
which is related to several string theory models
\cite{Stewart:1994ts,Cicoli:2008gp}, and is known as a class of
fibre inflation, in which case the potential is,
\begin{equation}\label{fibre}
V(\phi)=V_0^4 (1-\exp (-\lambda \kappa  \phi ))\, ,
\end{equation}
where again the constant $V_0$ has mass dimensions $[V_0]=[m]$ and
$\lambda$ is a dimensionless constant. In this case the slow-roll
index $\epsilon_1$ is,
\begin{equation}\label{epsilonalphaatr}
\epsilon_1=\frac{\alpha  \lambda ^2 e^{-2 \kappa  \lambda  \phi
}}{2 \left(1-e^{-\kappa  \lambda  \phi }\right)^2}\, ,
\end{equation}
and again by solving the equation $\epsilon_1(\phi_f)=1$, we
obtain $\phi_f$,
\begin{equation}\label{phifinal1}
\phi_f=\frac{\log \left(\frac{1}{2} \left(2-\sqrt{2} \sqrt{\alpha
} \lambda \right)\right)}{\kappa  \lambda }\, .
\end{equation}
Accordingly, by integrating (\ref{efoldingsnew}), we get $\phi_i$,
\begin{equation}\label{phii1}
\phi_i=\frac{\log \left(\frac{1}{2} \left(\sqrt{2} \sqrt{\alpha }
\lambda +2 \alpha  \lambda ^2 N+2\right)\right)}{\kappa  \lambda
}\, .
\end{equation}
Hence, the spectral index of the primordial scalar curvature
perturbations, and the tensor-to-scalar ratio at leading order in
$1/N$  are,
\begin{equation}\label{nsalphaattractor}
n_s\simeq 1-\frac{2}{N} -\frac{3}{\alpha  \lambda ^2
N^2}+\frac{\sqrt{2}}{\sqrt{\alpha } \lambda  N^2}\, ,
\end{equation}
\begin{equation}\label{tensoralphaattr}
r\simeq \frac{8}{\alpha  \lambda ^2 N^2}\, .
\end{equation}
In this case too, the model has good inflationary phenomenological
properties, and the viability can be achieved for a large range of
the free parameters $\alpha$ and $\lambda$, for $N\sim 60$.
However, as in the previous paradigm, we shall choose $\alpha$ and
$\lambda$ in such a way so that both the Swampland criteria are
satisfied and the inflationary phenomenology of the model is
guaranteed.
\begin{figure}
\centering
\includegraphics[width=18pc]{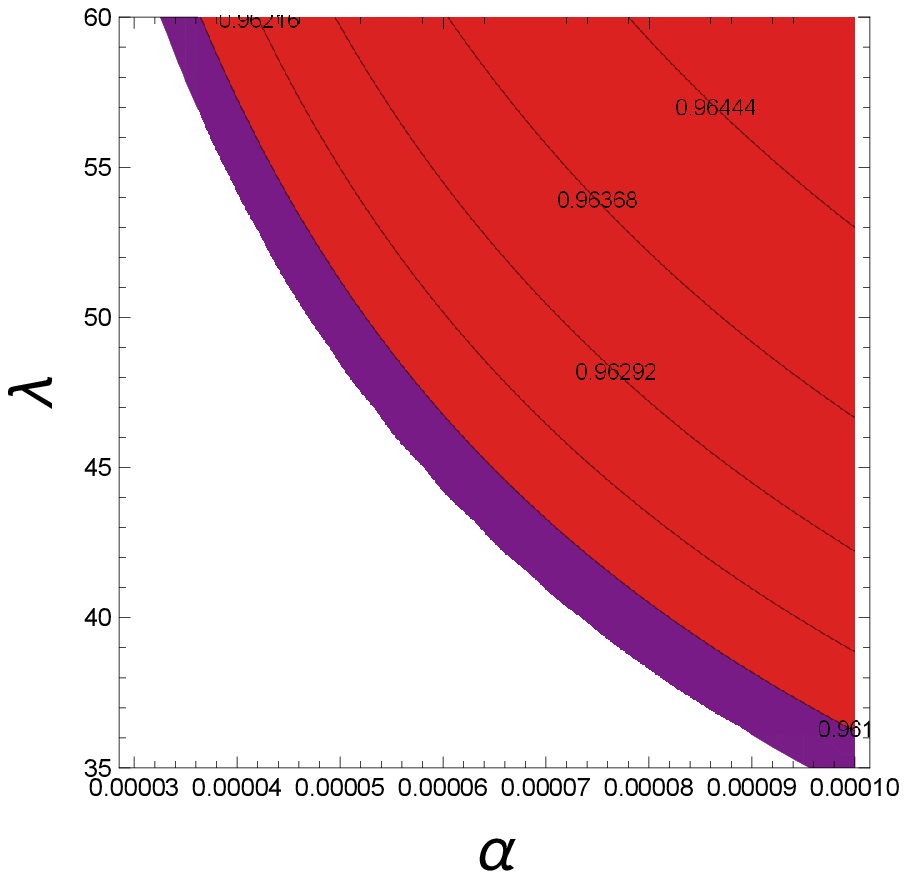}
\includegraphics[width=18pc]{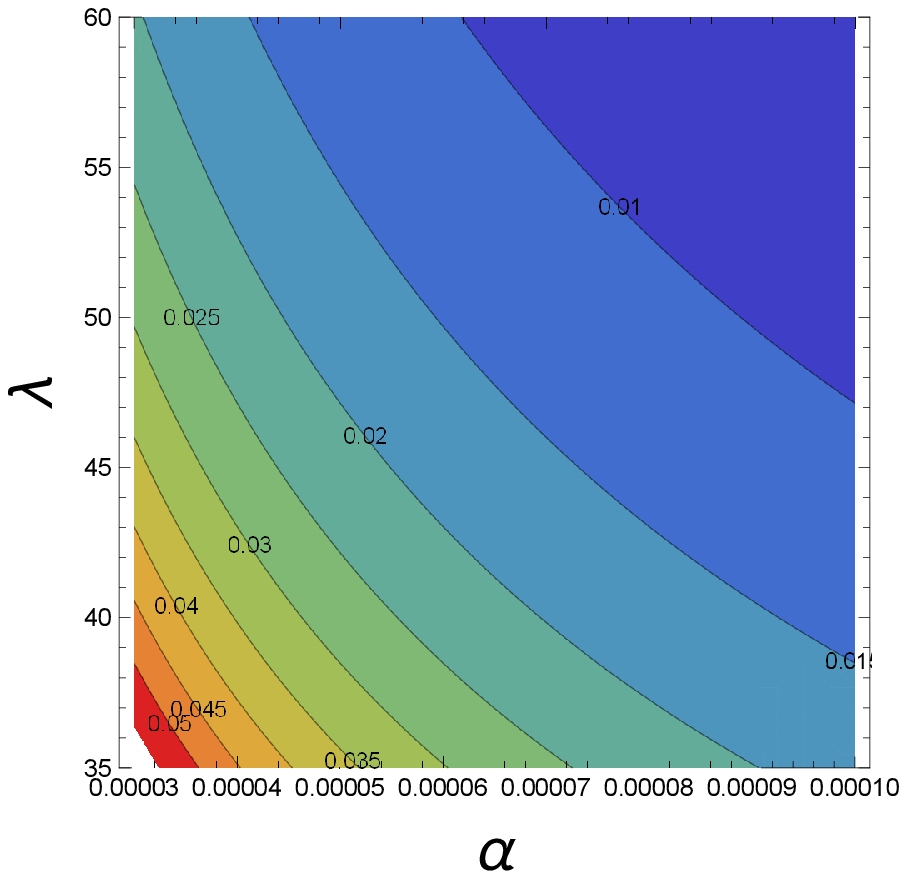}
\caption{Contour plots for the spectral index of primordial scalar
curvature perturbations $n_s$ (left plot) and the tensor-to-scalar
ratio $r$ (right plot), for $\lambda=[35,60]$ and
$\alpha=[0.00003, 0.0001]$ and for $N=60$.}\label{plot3}
\end{figure}
In Fig. \ref{plot3} we present the contour plots for the spectral
index of primordial scalar curvature perturbations values (left
plot) and the tensor-to-scalar ratio (right plot), for
$\lambda=[35,60]$ and $\alpha=[0.00003, 0.0001]$ and $N=60$, and
in both the plots of Fig. \ref{plot3}, we kept values in the plot
range that satisfy the latest Planck constraints, that is
$n_s=[0.9607, 0.9691]$ and $r=[0,0.056]$. As it can be seen, for
this model too, the simultaneous compatibility of both the
spectral index $n_s$ and of the tensor-to-scalar ratio $r$ is
achieved. Now let us turn our focus on the Swampland criteria, and
in Fig. \ref{plot4} we present the results of our numerical
analysis. Particularly, the left contour plot corresponds to the
values of $\frac{V'(\phi)}{V(\phi)}$, and also the right contour
plots corresponds to the values of $-\frac{V''(\phi)}{V(\phi)}$,
evaluated at the first horizon crossing for $\lambda=[35,60]$ and
$\alpha=[0.00003, 0.0001]$ and $N=60$. In the left plot we kept
the plot values of $\frac{V'(\phi)}{V(\phi)}$  in the range
$\frac{V'(\phi)}{V(\phi)}=[1,9]$ while for
$\frac{V''(\phi)}{V(\phi)}=[100,500]$, so both the quantities
$\frac{V'(\phi)}{V(\phi)}$ and of $-\frac{V''(\phi)}{V(\phi)}$
take values well above unity in reduced Planck units, especially
$-\frac{V''(\phi)}{V(\phi)}$ takes quite large values for the same
set of free parameters values .
Hence, in this case too, the last two Swampland criteria are
satisfied and also we assumed that $\Delta \phi \sim M_p$, or
$\Delta \phi \sim 1$ in reduced Planck units for the whole
analysis.

\subsection{Dark Energy Evolution for the $f(R)$ Gravity Model with
Fibre Inflation Potential}

\begin{figure}
\centering
\includegraphics[width=18pc]{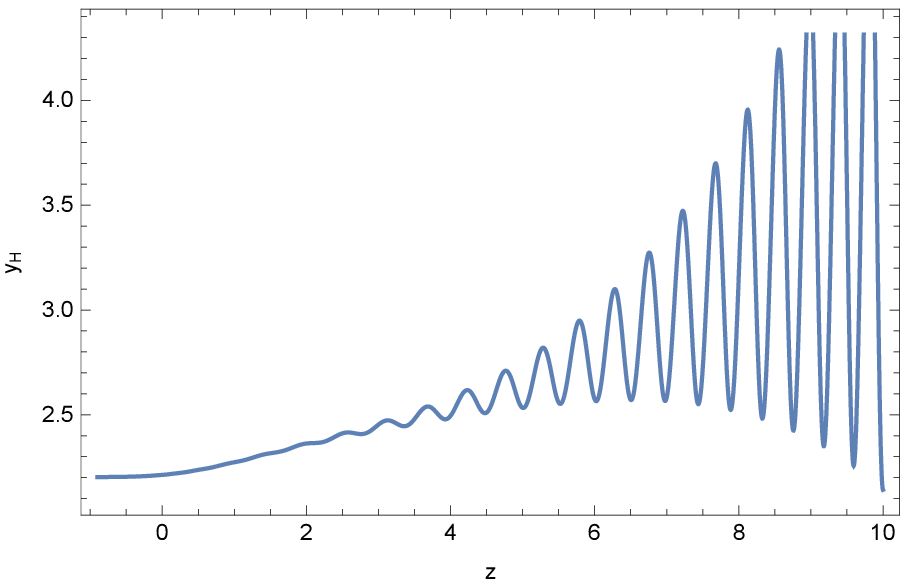}
\includegraphics[width=18pc]{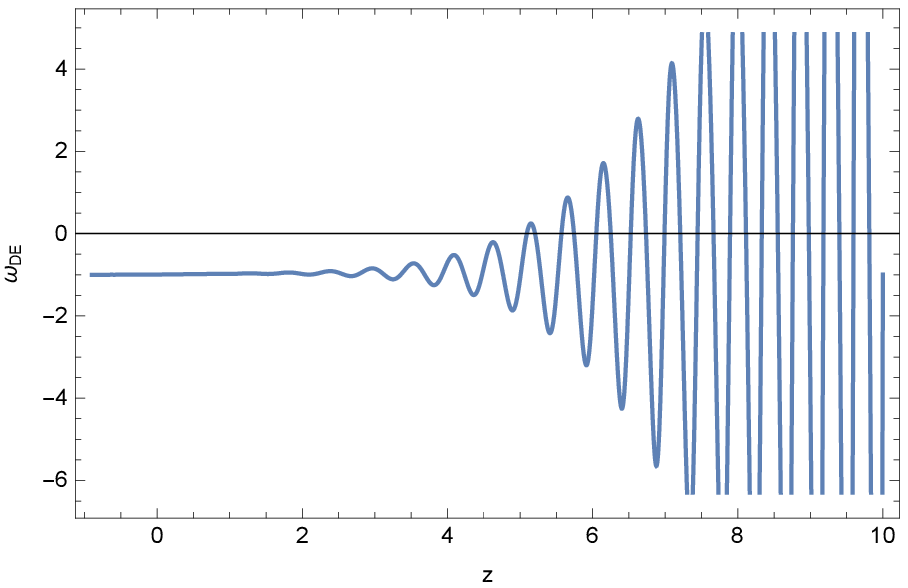}
\caption{The statefinder function $y_H$ for geometric dark energy
as a function of the redshift (left plot), and the dark energy EoS
parameter $\omega_{DE}$ (right plot).}\label{plot5}
\end{figure}

\begin{figure}
\centering
\includegraphics[width=18pc]{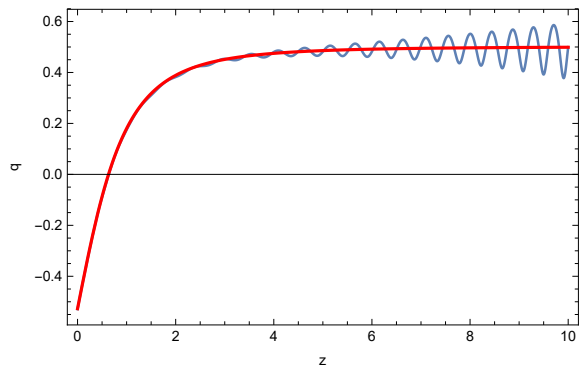}
\caption{The deceleration parameter, as a function of the redshift
for the fibre $f(R)$ gravity model (blue curve) and for the
$\Lambda$CDM model (red curve).} \label{plot6}
\end{figure}

The motivation for studying the exponential $f(R)$ gravity models,
was mainly the interesting late-time phenomenology these generate,
as is the case in Ref. \cite{Oikonomou:2020qah}, where the
unification of inflation with the late-time era was obtained. In
this work, we shall also consider the late-time phenomenology of
the model (\ref{action}) with the $f(R)$ gravity model being that
of Eq. (\ref{frini}) and the scalar potential of the fibre
inflation, namely that of Eq. (\ref{fibre}), in the presence of
cold dark matter and of radiation perfect fluids. The model of
this work and the one studied in Ref. \cite{Oikonomou:2020qah} are
different in their functional form, but have similarities,
however, in the present paper the inflationary era is controlled
by a canonical scalar field, while in the model of Ref.
\cite{Oikonomou:2020qah}, the early-time era was controlled by an
$R^2$ term, and the axion scalar was frozen in its primordial
vacuum expectation value. Let us now study the late-time behavior
of the $f(R)$ gravity canonical scalar field theory. To this end,
instead of the cosmic time we shall express all the dynamically
evolving quantities as functions of the redshift,
\begin{equation}
\centering \label{z} 1+z=\frac{1}{a(t)}\, ,
\end{equation}
and also we shall make extensive use of the following
differentiation rule,
\begin{equation}
\centering \label{dz} \frac{d}{dt}=-H(1+z)\frac{d}{dz}\, .
\end{equation}
More importantly, we shall express all the Hubble rate dependent
terms and their derivatives, as functions of the statefinder
function $y_H(z)$ and its derivatives
\cite{Hu:2007nk,Bamba:2012qi,Odintsov:2020qyw,Odintsov:2020nwm},
\begin{equation}
\centering \label{yH} y_H=\frac{\rho_{DE}}{\rho_m^{(0)}}\, ,
\end{equation}
with $\rho_{DE}$ standing for the dark matter energy density and
with $\rho_m^{(0)}$ being the current value of density for
non-relativistic matter. The forms of the dark energy density and
of the dark energy pressure can be derived by the field equations
corresponding to the action (\ref{action}),
\begin{equation}
\centering \label{DEdensity}
\rho_{DE}=\frac{1}{2}\dot\phi^2+V+\frac{f_R
R-f}{2\kappa^2}-\frac{3H\dot f_R}{\kappa^2}+\frac{3H^2}{\kappa^2}(1-f_R)\, ,
\end{equation}
\begin{equation}
\centering \label{PDE} P_{DE}=-V-\frac{f_R
R-f}{2\kappa^2}-\frac{2\dot H}{\kappa^2}(1-f_R)\, .
\end{equation}
The field equations can be cast in their Einstein-Hilbert form for
a FRW metric,
\begin{equation}
\centering \label{motion6}
\frac{3H^2}{\kappa^2}=\rho_{(m)}+\rho_{DE}\, ,
\end{equation}
\begin{equation}
\centering \label{motion7} -\frac{2\dot
H}{\kappa^2}=\rho_{(m)}+P_{(m)}+\rho_{DE}+P_{DE}\, ,
\end{equation}
where $\rho_{(m)}$ and $P_{(m)}$ are the energy density and the
pressure of the matter perfect fluids that are present, which we
shall assume that these are the radiation and cold-dark-matter
fluids. Hence, the Hubble rate can be written as a function of the
statefinder as follows,
\begin{equation}
\centering \label{H}
H^2=m_s^2\left(y_H(z)+\frac{\rho_{(m)}}{\rho_m^{(0)}}\right)\, ,
\end{equation}
where
$m_s^2=\kappa^2\frac{\rho_m^{(0)}}{3}=1.87101\cdot10^{-67}$eV$^2$.
We shall solve numerically the field equations in the redshift
interval $z=[0,10]$, by assuming the following initial conditions,
\begin{equation}\label{generalinitialconditions}
y_H(z_f)=\frac{\Lambda}{3m_s^2}\left(
1+\frac{(1+z_f)}{1000}\right)\, , \,\,\,\frac{d y_H(z)}{d
z}\Big{|}_{z=z_f}=\frac{1}{1000}\frac{\Lambda}{3m_s^2}\, ,
\end{equation}
for the function $y_H(z)$, where $z_f=10$, while for the scalar
field, the initial conditions are chosen as,
\begin{equation}\label{scalarfieldinitial}
\phi(z_f)=10^{-16}M_p,\,\,\, \phi'(z_f)=-10^{-7}M_p\, .
\end{equation}
As for the free parameters and the various physical parameters, we
shall assume that the Hubble rate at present time is $H_0\simeq
1.37187\times 10^{-33}$eV according to the latest Planck data
\cite{Aghanim:2018eyx}. In addition, $\Omega_{\Lambda}\simeq
0.681369$ and $\Omega_M\sim 0.3153$ \cite{Aghanim:2018eyx}, while
$\Omega_r/\Omega_M\simeq \chi$. Also for the numerical analysis of
the model (\ref{frini}), we shall choose $\gamma=6$, $\zeta=0.23$
$\delta=0.01$ and $\lambda=0.001$. In addition,
$m_s^2=\frac{\kappa^2\rho^{(0)}_m}{3}=H_0\Omega_c=1.37201\times
10^{-67}$eV$^2$, and $\Lambda$ is assumed to be equal to the
present time cosmological constant, while $V_0$ appearing in Eq.
(\ref{fibre}) is chosen to be $V_0=\Lambda^4$. In Fig. \ref{plot5}
we present the statefinder $y_H(z)$ as a function of the redshift
(left plot) and the dark energy EoS parameter as a function of the
redshift (right plot).
It is apparent that the dark energy oscillations are present even
in this scalar $f(R)$ gravity model, while in the model of Ref.
\cite{Oikonomou:2020qah}, dark energy oscillations were absent.
Hence it seems that the dark energy oscillations seem to be a
model dependent feature of modified gravity models. In order to
have a concrete quantitative idea of how viable is the model under
study, let us calculate the dark energy equation of state (EoS)
parameter $\omega_{DE}$, and the dark energy density parameter
$\Omega_{DE}$ at present time, for the model under study. In terms
of $y_H$, the aforementioned quantities are defined as
\cite{Bamba:2012qi,Odintsov:2020qyw,Odintsov:2020nwm},
\begin{align}
\centering \label{DE}
\omega_{DE}&=-1+\frac{1+z}{3}\frac{d\ln{y_H}}{dz}&\Omega_{DE}&=\frac{y_H}{y_H+\frac{\rho_{(m)}}{\rho_m^{(0)}}}\,
,
\end{align}
so our numerical analysis yielded the following values,
\begin{equation}
\centering \label{DE}
\omega_{DE}(0)=-0.996093,\,\,\,\Omega_{DE}(0)=0.688719\, ,
\end{equation}
which are compatible with the latest Planck constraints on the
cosmological parameters \cite{Aghanim:2018eyx}, which indicate
that $\omega_{DE}=-1.018\pm 0.031$, and $\Omega_{DE}=0.6847\pm
0.0073$. Finally, it is worth to compare in brief the present
model with the $\Lambda$CDM model, so let us consider a
statefinder quantity in order to reveal the resemblance of the
$\Lambda$CDM model with the fibre $f(R)$ gravity model. We shall
consider the deceleration parameter,
\begin{align}\label{statefinders}
& q=-1-\frac{\dot{H}}{H^2}=-1+(z+1)\frac{H'(z)}{H(z)}\, ,
\end{align}
and in Fig. \ref{plot6} we present the deceleration parameter for
the fibre $f(R)$ model (blue curve), with the $\Lambda$CDM
deceleration parameter (red curve). As it can be seen in Fig.
\ref{plot6} the two models are indistinguishable for $z<4$. Also
at present day, the deceleration parameter for the fibre $f(R)$
gravity model is $q(0)=-0.528993$, while the $\Lambda$CDM value is
$q=-0.527$, which are quite similar values.
Finally, our numerical analysis indicated that the initial
condition of the scalar field at $z=z_f$ does not affect
significantly the late-time qualitative behavior, however, the
initial condition of the derivative of the scalar field with
respect to the redshift variable, affects significantly the
late-time behavior.

In conclusion, we demonstrated that the $f(R)$ gravity scalar
field model with fibre inflation potential, can describe in an
unified way both the inflationary era and the dark energy era,
producing results which are both compatible with the 2018 Planck
constraints on inflation \cite{Akrami:2018odb} and with the 2018
Planck constraints on the cosmological parameters
\cite{Aghanim:2018eyx}. At the same time the model formally
satisfies the Swampland criteria, for the same set of values of
the free parameters that yield the inflationary viability of the
model. The new result that the fibre $f(R)$ gravity model brings
along is that the early-time era is essentially governed by a
rescaled Einstein-Hilbert canonical scalar theory, and the rest of
the $f(R)$ terms affect only the late-time era. The presence of a
rescaled Ricci scalar term is a solid effect of modified gravity,
which affects the inflationary effective Lagrangian at large
curvatures.

Finally it is worthy discussing which term affects the most the
dark energy era, is it the $f(R)$ gravity or the scalar field. The
answer to this question can be given only qualitatively, and it
seems that only $f(R)$ gravity drives the late-time era. The only
effect of the scalar field is that it mainly affects the dark
energy oscillations, which, in the absence of the scalar field and
for the class of $f(R)$ models studied in this paper, are absent,
see for example Ref. \cite{Oikonomou:2020qah}. Also the viability
of the model is obtained for different values of the exponent of
the power-law $f(R)$ gravity term. Thus the scalar model changes
the behavior of the dark energy era in a qualitative way since it
changes the parameter range for which viability can occur, and
also it re-introduces oscillations in the class of models studied
in the present paper, which were absent in the scalar field free
theory of Ref. \cite{Oikonomou:2020qah}.

\section{Conclusions}

In this work we presented a model of $f(R)$ gravity in the
presence of a canonical scalar field, and we investigated a
unification of inflation with dark energy scenario. The $f(R)$
gravity was chosen to be of exponential type with power-law
corrections with their exponent being less than unity and
positive. During the large curvature era, the $f(R)$ gravity at
leading order became a rescaled type of Einstein-Hilbert gravity,
with the dominant term being of the form $\alpha R$, with $\alpha$
being a dimensionless constant. Thus the effective Lagrangian
during the inflationary era was that of a rescaled
Einstein-Hilbert canonical scalar theory. For the inflationary era
we investigated the phenomenologically viability of the resulting
theory and also we examined whether it was possible to comply with
the Swampland criteria. As we showed, this is possible for the
resulting theory, if the first Swampland criterion holds
marginally true, that is, if the scalar field during the
inflationary era is of the order $\phi\sim M_p$. We exemplified
our study by using two characteristic inflationary potentials, one
supergravity related $\alpha$-attractors model, and one fibre
inflation model. After examining carefully the parameter space of
the two models, we showed that the phenomenologically viability of
the models is easy to achieve, and also we demonstrated that for
the same set of values for the free parameters that guarantee the
inflationary viability of the models, the models became compatible
with the Swampland criteria. Finally, we addressed the question
whether it is possible to describe the inflationary and the dark
energy era within this theoretical framework of $f(R)$ canonical
scalar theory. By investigating the fibre inflationary potential
at late times, using a numerical approach, we demonstrated that
the $f(R)$ scalar theory can produce a viable dark energy, which
is compatible with the latest 2018 Planck data and mimics the
$\Lambda$CDM model, when statefinder quantities are considered. A
notable feature characterizing the dark energy era, is the
presence of dark energy oscillations, and by taking into account
the results of Ref. \cite{Oikonomou:2020qah}, we may reach the
conclusion that the dark energy oscillations feature is a model
depended feature in modified gravity dark energy models.

\end{document}